\documentclass[epj]{svjour}
\usepackage{amsmath,epsfig}

\usepackage{graphicx}
\usepackage{fancyhdr}
\def\makeheadbox{{
 }}
\def\mytitle{My title} 
\def\myauthors{My name}  
\def\mytype{My type of session}
\def\mysession{My session}

\def\mytitle{$(g-2)_\mu$ and supersymmetry: status and prospects} 
\def\myauthors{Dominik St\"ockinger}    
\def\mytype{Contributed Talk}    
\def\mysession{Flavor Physics}

\def\amu{a_\mu}

\begin{document}
\title{\boldmath{$(g-2)_\mu$} and supersymmetry: status and prospects}

\author{Dominik St\"ockinger\inst{}
\thanks{\emph{Email:}  d.stockinger@physics.gla.ac.uk}%
}                     
%
%
\institute{Department for Physics \& Astronomy,
University of Glasgow}
%
\date{}
\abstract{
The experimental determination of the muon magnetic moment and its
theoretical prediction within the Standard Model and the MSSM are
reviewed. A 3$\sigma$ deviation between experiment and Standard Model
prediction has been established, and supersymmetry could provide a
natural explation of this deviation. Possible future improvements and
the case for a new experiment are discussed.
} 
\maketitle
%

\section{Introduction}
\label{intro}

For decades, the main arguments in favour of supersymmetry at or below
the TeV-scale have been of a theoretical nature and have been related to
e.g.\ naturalness in the Higgs sector or unification of gauge
couplings. Now, the muon magnetic moment
$\amu=(g-2)_\mu/2$ has developed into one of the strongest and most
robust observational indications for the 
existence of supersymmetry at or below the TeV-scale. Independent of
theoretical naturalness or unification arguments, a 3$\sigma$ deviation
between the experimental and Standard Model (SM) theory value of $\amu$
has been established, and this deviation can be well explained by
TeV-scale 
supersymmetry but is very hard to accomodate within many other scenarios
for physics beyond the SM. In these proceedings\footnote{These
proceedings are based on
\cite{review,WhitePaper}.} we briefly review the
current status of $\amu$ within the SM and the minimal supersymmetric
Standard Model (MSSM) and we discuss possible future improvements.

\section{Deviation between the experimental value and the SM theory
 prediction}
\label{sec:1}

The muon magnetic moment has been measured at the recent E821 experiment
at Brookhaven. The final result of this experiment reads \cite{BNLfinal}
\begin{align}
\amu^{\rm exp} &= 11\,659\,208.0(6.3)\times10^{-10}.
\end{align}
The success of this experiment has inspired tremendous progress also on
the SM theory evaluation of $\amu$, particularly on the hadronic
vacuum polarization contributions and hadronic light-by-light
contributions, the two contributions with the by far largest theory
uncertainties.

The hadronic light-by-light contributions are tiny but important at the
current level of precision. They cannot be evaluated from first
principles. In the 1990's these contributions have been evaluated by two
groups \cite{BPP,HayakawaKinoshita}. Since then, major 
progress has been made in two directions: first a sign error in these
calculations was uncovered in \cite{KnechtNyffeler} (and confirmed by the
original authors), the correction of which shifted the SM theory
prediction by more than $+16\times10^{-10}$. Second, new short-distance
constraints on the relevant light-by-light correlator were studied and
incorporated into the computation in \cite{MelnV}, which again shifted
the result by about $5\times10^{-10}$. The recent developments are
reviewed in more detail in
\cite{BijnensPrades,MillerRR,CzarneckiProc}, and current
estimates for the hadronic light-by-light contributions vary between
\begin{align}
\amu^{\rm
LbL}&=10.0(3.9)\mbox{\cite{Jegerlehner}}\ldots13.6(2.5)\mbox{\cite{MelnV}}.
\end{align}

The hadronic vacuum polarization contributions are currently the
dominant source of the SM theory uncertainty. Via the optical theorem,
they can be related to the cross section for $e^+e^-\to$ hadrons, which can
be measured. The recent progress is due to refined ways to combine
existing experimental data on $e^+e^-\to$ hadrons, obtained from
different experiments and for different energies, and to improved
measurements of $e^+e^-\to$ hadrons. In the last 10 years, new results
on this cross section have become available from BES-II \cite{BES},
CMD-2 \cite{CMD2a}, and most recently from SND \cite{SND} and CMD-2
\cite{CMD2b}, both in Novosibirsk, and from KLOE \cite{KLOE} and
BaBar. The KLOE measurement is particularly interesting since it is the
first one using radiative return measurements. Three major groups
\cite{Davier06,Teubner06,Jegerlehner} have presented updated
evaluations that incorporate the latest measurements\footnote{They
differ e.g.\ in the way they incorporate the KLOE data.}, with 
results in the range 
\begin{align}
\amu^{\rm vac.pol.}&=
689.4(4.6) \mbox{\cite{Teubner06}}\ldots 692.1(5.6) \mbox{\cite{Jegerlehner}}.
\end{align}
In principle, part of the $e^+e^-\to$ hadrons cross section could be
obtained in an alternative way from hadronic $\tau$ decays
\cite{Alemanyetal}. This was particularly useful when the $e^+e^-$ data
was rather imprecise and dominated by only the CMD-2 data. Now, several
$e^+e^-$ data sets are available, and a disagreement between $e^+e^-$
based and $\tau$ data based analyses of $\amu^{\rm vac.pol.}$ has led
most groups to a preference of the theoretically cleaner $e^+e^-$ based
analyses (see e.g.\ the discussions in \cite{Davier06,MillerRR}).

After the most recent progress the SM theory prediction for $\amu$ has
reached a very mature state. The full prediction is obtained by adding
the QED and electroweak to the hadronic contributions. The review
\cite{MillerRR} obtains
\begin{align}
\amu^{\rm SM} &=   11\,659\,178.5(6.1)\times10^{-10}
\label{SMpred}
\end{align}
and thus
\begin{align}
\amu^{\rm exp}- \amu^{\rm SM} &=29.5(8.8)\times10^{-10},
\label{deviation}
\end{align}
a $3.4\sigma$ deviation! The results obtained in
\cite{Davier06,Teubner06,Jegerlehner} differ slightly but all obtain
deviations of more than $3\sigma$. Therefore a 3$\sigma$ deviation
between the experimental and the SM theory value of $\amu$ has been
firmly established.

\section{Muon magnetic moment and supersymmetry}

If the observed $3\sigma$ deviation is not due to an error or a
statistical fluctuation, where could it come from? If supersymmetry
(SUSY) exists, the superpartner particles would give rise to a
contribution to $\amu$ of approximately
\begin{align}
\amu^{\rm SUSY} &\approx  13\times10^{-10}\left(\frac{100\,\rm
    GeV}{M_{\rm SUSY}}\right) ^2\ \tan\beta\ \mbox{sign}(\mu),
\label{SUSYapprox}
\end{align}
where $M_{\rm SUSY}$ denotes the common superpartner mass scale,
$\tan\beta$ the ratio of the two Higgs vacuum expectation values, and
$\mu$ the Higgsino mass parameter. Hence, supersymmetry could easily
be the origin of the observed deviation of $29.5\times10^{-10}$, e.g.\
for SUSY masses of roughly 200 GeV and $\tan\beta\sim10$ or
SUSY masses of 500 GeV and $\tan\beta\sim50$.

Although this result is very well known and has been stressed many
times, see e.g.\ \cite{CzMarciano}, it is quite non-trivial and
singles out supersymmetry among many extensions of the SM. One should
note that the deviation of $29.5\times10^{-10}$ is almost twice as
high as the SM electroweak contributions, i.e.\ diagrams with $W$,
$Z$, Higgs exchange etc.,
\begin{align}
\amu^{\rm SM\ EW} &= 15.4(0.2)\times10^{-10}.
\end{align}
Likewise, a generic extension of the SM with weakly interacting
particles and  characteristic mass scale $M_{\rm BSM}$  will be
suppressed by $(M_W/M_{\rm BSM})^2$ and lead to
contributions of the order
\begin{align}
\amu^{\rm BSM\ generic} &\propto \left(\frac{300\mbox{
      GeV}}{M_{\rm BSM}}\right)^2\times10^{-10},
\end{align}
which is far too small except for very small $M_{\rm BSM}$, which is
typically already ruled out.

\subsection{$\tan\beta$ enhancement}

Supersymmetry has two advantages compared to such generic extensions
of the SM: First, masses for the relevant supersymmetric particles,
mainly smuons and charginos as small as $M_{\rm SUSY}\sim100$~GeV are 
still experimentally allowed. Second, the parameter $\tan\beta$ can 
provide an enhancement by a factor of up to about 50.

The $\tan\beta\ \mbox{sign}(\mu)$ behaviour can be easily explained on a
diagrammatic level. Each diagram contributing to $\amu$ must contain a
chirality flip between a left- and a right-handed (s)muon.  The
$\tan\beta$-enhancement arises in diagrams where the 
necessary chirality flip occurs at a muon Yukawa coupling, either to a
Higgsino or Higgs boson, because this coupling is enhanced by
$1/\cos\beta\approx\tan\beta$ compared to its SM value. The
$\mu$-parameter mediates the transition between the two Higgs/Higgsino
doublets $H_{1,2}$, and this transition enhances diagrams because only
$H_1$ couples to muons while $H_2$ has the larger vacuum expectation
value, $v_2/v_1=\tan\beta$. Therefore, all $\tan\beta$-enhanced terms
are also proportional to sign($\mu$). This behaviour is not restricted
to the one-loop level but repeats itself in higher orders.

\subsection{Status of the MSSM prediction for the muon magnetic moment}

The fact that supersymmetry is potentially the origin of the observed
$3\sigma$ deviation justifies a precise analysis of the prediction for
$\amu$ within the MSSM (for a review see \cite{review}). The MSSM
prediction is given by the SM prediction plus the genuine SUSY
contributions, arising from diagrams with SUSY particle loops.

The SUSY
one-loop contributions consist of diagrams with chargino/sneutrino or
neutralino/smuon loops. These diagrams have been known for a long time
\cite{oneloop}. The full expression is not repeated here. The
approximation (\ref{SUSYapprox}) can serve as a guideline. The mass
parameters governing the one-loop SUSY contributions are mainly the 
left-handed smuon mass $m_{L,\tilde{\mu}}$ and the gaugino mass $M_2$,
while $\mu$ and the right-handed smuon mass $m_{R,\tilde{\mu}}$ have a
smaller influence. If all mass parameters are equal to $M_{\rm
SUSY}$, (\ref{SUSYapprox}) is an excellent approximation. If there are
mass splittings, (\ref{SUSYapprox}) still provides a reasonable estimate
if $M_{\rm SUSY}$ is identified with a value between $m_{L,\tilde{\mu}}$
and $M_2$. Contrary to the other mass parameters, increasing $\mu$ can
lead to enhancements, e.g.\ if $m_{L,\tilde{\mu}}\approx
m_{R,\tilde{\mu}}\approx M_2\ll\mu$ due to a diagram with bino exchange,
which is directly $\propto\mu$.

At the two-loop level two kinds of SUSY contributions are
known. QED-logarithms  $\log(M_{\rm SUSY}/m_\mu)$ arising from SUSY
one-loop diagrams with additional photon exchange have been evaluated in
\cite{twoloopB} and amount to $-7\%\ldots-9\%$ of the one-loop
contributions. Two-loop diagrams involving closed loops of either
sfermions (stops, sbottoms, etc) or charginos/neutralinos have been
evaluated in \cite{twoloopA}. They amount to about $2\%$ of the one-loop
contributions if all SUSY masses are degenerate but can be much larger,
e.g.\ if smuon masses are very heavy but stops and/or 
charginos and Higgs bosons are light.

A very important question regards the remaining theory error of the SUSY
contributions to $\amu$. This theory error arises from unknown two-loop
and higher order contributions. It has been estimated in 
\cite{review} to 
\begin{eqnarray}
\delta\amu^{\rm SUSY}(\mbox{unknown}) &= 0.02\ \amu^{\rm SUSY,1L} +
 2.5\times10^{-10},\  \ 
\label{amuSUSYerror}
\end{eqnarray}
which is smaller than the current SM theory error and the experimental
uncertainty. 

\subsection{Implications on SUSY phenomenology}

\begin{figure}
\begin{picture}(5,130)(0,2.)
\epsfysize=5.0cm\put(-0.2,3.2){\epsfbox{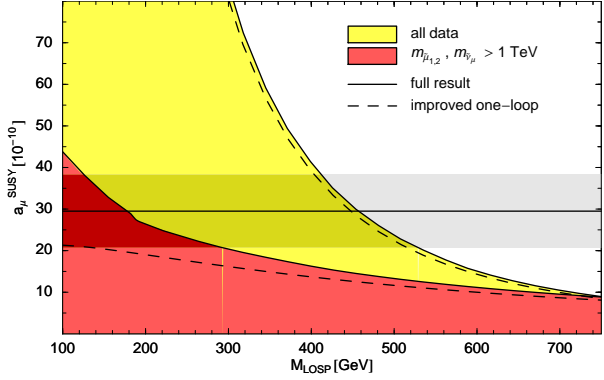}}
\end{picture}
\caption{Possible values of $\amu^{\rm SUSY}$ as a function of the mass
 of the lightest observable SUSY particle $M_{\rm LOSP}$, from a scan of
 the MSSM parameter space and for $\tan\beta=50$. The light yellow
 region corresponds to all data points; the red region corresponds to
 points with smuons and sneutrinos that are heavier than 1 TeV. The
 deviation (\ref{deviation}) is also indicated.}
\label{fig:1}       
\end{figure}

Fig.\ \ref{fig:1} summarizes the current status of $\amu$ and SUSY.  A
scan of the MSSM parameter space 
has been performed (for $\tan\beta=50$ and taking into account
experimental constraints from e.g.\ Higgs searches and $b$-physics; for
further details see \cite{review}), and the resulting values for
$\amu^{\rm SUSY}$, including all known one- and two-loop contributions,
are plotted as a function of the mass of the lightest observable SUSY
particle. Fig.\ \ref{fig:1}
confirms again that SUSY can easily explain the observed deviation if
$M_{\rm LOSP}$ is below about 600 GeV. 

Apart from $\amu$, significant information on SUSY parameters can be
inferred from the measured dark matter density, if it is assumed to
consist of the stable, lightest SUSY particle. The two observables tend
to constrain orthogonal directions in the multi-dimensional SUSY
parameter space and are thus complementary. Several recent comprehensive
studies \cite{Rosk,All,Ellis,Sfitter} have shown that the MSSM is able to
simultaneously accomodate all existing data from $\amu$, dark matter,
$b$-physics and electroweak precision observables. This is even
possible, in spite of some slight tensions, in the constrained MSSM
(CMSSM), a model with only 4 input parameters. One result of these
studies is that rather low SUSY masses are preferred as a consequence
of the $\amu$ deviation.

\section{Future prospects and the case for a new experiment}

The present $3\sigma$ deviation is one of the strongest observational
hints for the existence of supersymmetry at or below the
TeV-scale. However, although the deviation is tantalizing it is not
quite large enough to be regarded as a proof of physics beyond the
SM. Fortunately, there are good prospects that the current uncertainty
of $8.8\times10^{-10}$ of the deviation (\ref{deviation}) can be reduced
significantly in the near future.

\begin{figure}
\begin{picture}(5,130)(0,2)
\epsfysize=5cm\put(-0.2,3.2){\epsfbox{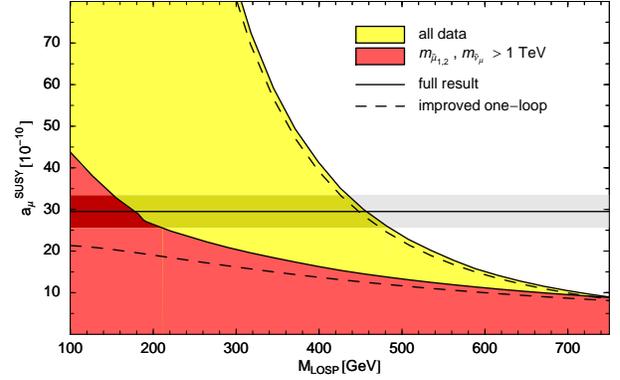}}
\end{picture}
\caption{As fig.\ \ref{fig:1} but showing the future precision of the
 deviation (\ref{future}).}
\label{fig:2}       
\end{figure}

The current theory error of $6.1\times10^{-10}$ of the SM prediction
(\ref{SMpred}) will soon decrease due to currently ongoing more precise
determinations of the $e^+e^-\to$ hadrons cross section. Both KLOE and
BaBar will soon release data on the most important $\pi\pi$ channel
using radiative return measurements. If these data are in agreement with
the Novosibirsk data, they will not only reduce the error but also
significantly increase our confidence in the $e^+e^-$ data. The new
data will immediately improve our knowledge of the hadronic vacuum 
polarization contributions to $\amu$, which currently are the dominant
source of error. 

The second most important source of theory error are the hadronic
light-by-light contributions. These are notoriously difficult to
evaluate, but they have moved into the centre of attention, and several
groups are currently investigating these contributions, using both
established and novel approaches. A determination of these
contributions with a relative accuracy of about $15\%$ seems
possible. In combination, a reduction of the theory error of the SM
prediction down to $(3\ldots5)\times10^{-10}$ within the next few years
seems likely.

The tantalizing status of the current deviation, together with the
prospect for an improvement of the SM theory prediction, highlights the
need for and the potential of a new, better experimental measurement of
$\amu$ \cite{WhitePaper}. A corresponding experiment, E969 at Brookhaven
\cite{E969}, with 
the goal of a final uncertainty of $2.5\times10^{-10}$, has been
proposed and received scientific approval at Brookhaven.

\begin{figure}
\begin{picture}(5,130)
\epsfysize=5cm\put(40.2,3.2){\epsfbox{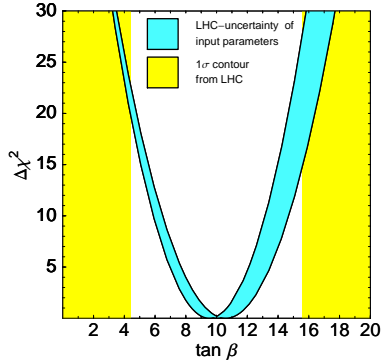}}
\end{picture}
\caption{Determination of $\tan\beta$ from LHC \cite{Sfitter} (yellow
 region) and from $\amu$ (blue band), 
 assuming (\ref{future}).}
\label{fig:3}       
\end{figure}

In an optimistic scenario, where the theory error is reduced to
$3\times10^{-10}$ and the magnitude of the deviation between SM theory
and experiment remains the same, this new measurement would lead to
\begin{align}
\amu^{\rm exp}- \amu^{\rm SM}(\mbox{future}) &=29.5(3.9)\times10^{-10}.
\label{future}
\end{align}
This more than 7$\sigma$ deviation would  dramatically sharpen the case
for new physics. The impact it would have on SUSY phenomenology
is illustrated in figs.\ \ref{fig:2}, \ref{fig:3}
\cite{WhitePaper}. Fig.\ \ref{fig:2} 
shows the same scan of the possible SUSY contributions to $\amu$ as in
fig.\ \ref{fig:1}, versus the future deviation. The precision of 
(\ref{future}) would lead to strong upper and lower mass bounds on SUSY
particles which could complement mass measurements from LHC.

Fig.\ \ref{fig:3} illustrates how $\amu$ might complement even
comprehensive LHC measurements. The analysis in
\cite{SfitterOld,Sfitter} shows that using a global fit of the MSSM to
LHC data one can determine SUSY masses rather precisely but the
parameter $\tan\beta$ rather poorly. If the benchmark point SPS1a
\cite{SPS} is realized, the LHC-analysis of \cite{SfitterOld} yields
$\tan\beta=10.22\pm9.1$, the improved analysis of \cite{Sfitter} yields
$\tan\beta=10\pm4.5$. Since $\amu^{\rm SUSY}$ is directly proportional
to $\tan\beta$, a precise determination as in (\ref{future}) would
provide an invaluable complement to LHC in the determination of
$\tan\beta$. Fig.\ \ref{fig:3} shows the value of
$\Delta\chi^2=(\amu^{\rm SUSY+SM}-\amu^{\rm
exp})^2/(3.9\times10^{-10})^2$ as a function of $\tan\beta$. In
$\amu^{\rm SUSY}$ all parameters except for $\tan\beta$ have been fixed
to the SPS1a values, which are accessible well at the LHC.

\section{Conclusions}

A tantalizing deviation of more than $3\sigma$ between the SM theory
prediction and the experimental value of $\amu$ has been
established. Supersymmetry with rather light masses and moderate to
large $\tan\beta$ could easily be the origin of this deviation. The near
future is very promising if the proposed E969 experiment \cite{E969} is
realized. The SM theory uncertainty will soon further
decrease and a new experiment could push the significance of the
deviation up to more than 7$\sigma$.

{\em Acknowledgments:} It is a pleasure to thank the organizers of
SUSY07 for this enjoyable conference.

%

%
%

\end{document}